# Quantum Spin Correlation Amplification Enables Macroscopic Detection of Atomic-Level Fatigue in Ferromagnetic Metals

Benniu Zhang[1,*,†], Liangshuo Zhang[1,†], Xiaodong Wu[1,†], Jigang Yu[1,†], Xiaochuan Cao[1], Zhijian Zhang[1], Xin Li[1], Fupeng Zhou[1], Jinglin Pan[1], Haifei Jiang[1], Gang Zheng[1,*]

[1]Chongqing Jiaotong University, Chongqing, 400074, China

[†]These authors contributed equally to this work

*Corresponding author. Email: benniuzhang@cqjtu.edu.cn; zhenggang@cqjtu.edu.cn

**Abstract:** Structural fatigue failures account for most of catastrophic metal component failures, annually causing thousands of accidents, tens of thousands of casualties, and $100 billion in global economic losses. Current detection methods struggle to identify early-stage fatigue damage characterized by sub-nanometer atomic displacements and localized bond rupture. Here we present a quantum-enhanced monitoring framework leveraging the fundamental symbiosis between metallic bonding forces and magnetic interactions. Through magnetic excitation of quantum spin correlation in metallic structures, we establish a macroscopic quantum spin correlation amplification technology that visualizes fatigue-induced magnetic flux variations corresponding to bond strength degradation. Our multi-scale analysis integrates fatigue life prediction with quantum mechanical parameters (bonding force constants, crystal orbital overlap population) and ferromagnetic element dynamics, achieving unprecedented prediction accuracy ($R^2 > 0.9$, $p < 0.0001$). In comprehensive fatigue trials encompassing 193 ferromagnetic metal specimens across 3,700 testing hours, this quantum magnetic signature consistently provided macroscopic fracture warnings prior to failure - a critical advance enabling 100% early detection success. This transformative framework establishes the first operational platform for preemptive fatigue mitigation in critical infrastructure, offering a paradigm shift from post-failure analysis to quantum-enabled predictive maintenance.

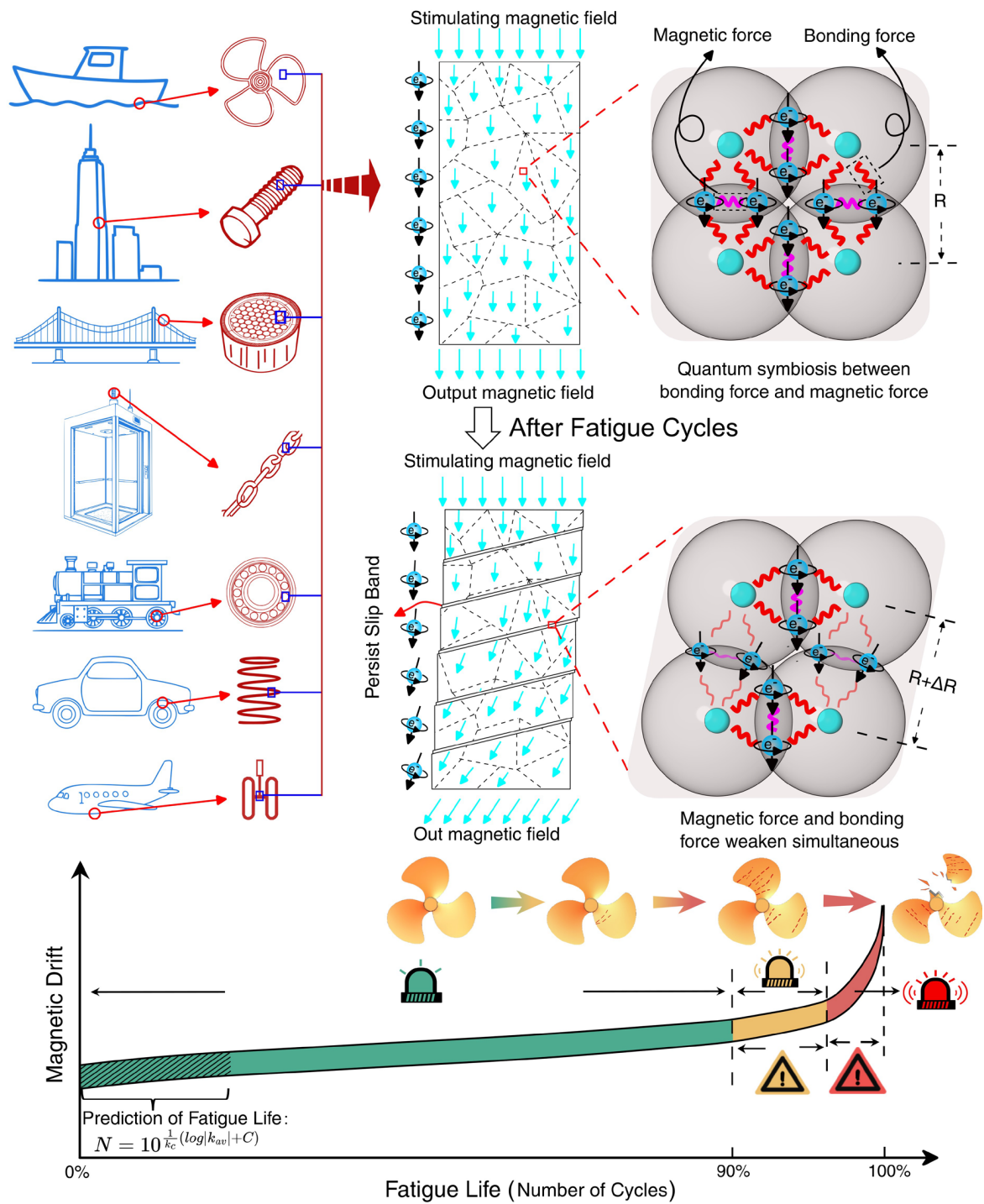

**Fig. 1 Graphical Abstract.** Regarding the issue of fatigue fracture in various industries. Leveraging the quantum symbiosis between atomic binding forces and magnetic interactions, we pioneer a quantum spin correlation approach via external magnetic excitation. It achieves accurate prediction of fatigue life and enables precise early warning prior to fracture onset, successfully identifying precursors of catastrophic failure before structural breakdown occurs, ensuring full prevention of fatigue-related structural disasters.

**Introduction**

Fatigue fracture is a key cause of metal structural failure[1-4], and the fatigue failure of actual structures often occurs abruptly and without warning. In many critical areas that are vital to the daily lives of the public, especially for large structures such as airplanes, ships, trains, bridges, elevators, and buildings, fatigue accidents occur frequently, with thousands of incidents each year, resulting in tens of thousands of casualties. At the same time, the costs associated with fatigue maintenance amount to nearly 100 billion US dollars annually (Fig. 2c and Methods).

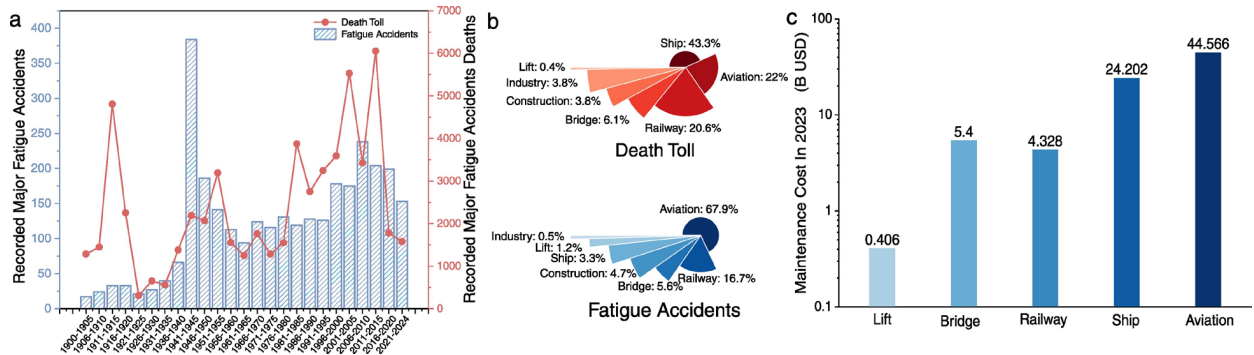

**Fig. 2 Incomplete statistics on recorded major fatigue accidents and deaths of large structures from 1900 to 2024, as well as costs related to fatigue maintenance of five industries in 2023. a**, The number of metal fatigue accidents and fatalities that occur every five years. **b**, The proportion of metal fatigue accidents and fatalities in different industries. **c**, Costs related to fatigue maintenance for the five large industries in 2023.

Current research predominantly concentrates on the micro-origins and evolution of fatigue, the temperature entropy evolution theory in fatigue, and S-N curve-based fatigue life predictions.

Microscopic investigations of fatigue phenomena primarily depend on observational techniques such as metallographic microscopes, Scanning Electron Microscopy (SEM), Transmission Electron Microscopy (TEM), and High-Resolution Digital Image Correlation (HR-DIC). These technologies enable the capture of image details at the nanometer scale, offering a novel perspective on the origins and evolution mechanisms of fatigue damage. Studies have identified a correlation between slip amplitude, the irreversibility of material behavior, and fatigue life, suggesting that the fatigue strength of metallic materials can be forecasted by monitoring the extent of slip localization during loading cycles[3]. In-situ micro-fatigue experiments concentrate on Persistent Slip Band (PSB) behavior at the micrometer scale[4]. Through meticulous observation and analysis, the formation and evolution mechanisms of PSBs, as well as the intricate processes of PSB nucleation and development within metals, have been elucidated.

Research into temperature entropy evolution in fatigue examines the relationship between material energy dissipation and fatigue damage. The theoretical foundation is that approximately 90% of the energy dissipated during fatigue loading is converted into thermal energy, leading to changes in the material's surface temperature[5]. This temperature variation is linked to the microstructural evolution of the material and the accumulation of fatigue damage[5]. Most studies concentrate on the diffusion characteristics of infrared thermography heat sources, fatigue life prediction models[6], and failure criteria.

The S-N curve, which plots stress or strain against fatigue life N, is a correlation curve established through experimental data[7,8]. Despite being a century and a half old, it remains the cornerstone of fatigue strength design and an essential component of material fatigue life assessment.

The aforementioned studies can be broadly categorized into two approaches: one that captures surface image data during the fatigue process for phenomenological analysis, and the other that extracts empirical formulas to describe phenomena through the fitting of experimental data. Regrettably, existing research has not yet explored the core intrinsic mechanism of fatigue fracture, which involves the interaction between atomic and electron clouds in materials. Moreover, the current evaluation of fatigue states in real structures is highly unsatisfactory, with prediction errors sometimes exceeding an order of magnitude[9].

**The essence of fatigue fracture: interatomic bonding gradually weakens until less than loading force**

The microscopic essence of metal material fracture is the disruption of metallic bonds within intact crystals. Consequently, its theoretical fracture strength is determined by the fracture limit of these metallic bonds[10]. Metal bonds are formed by the electrostatic coulomb force, or interatomic bonding force, between the electron cloud of atoms that have lost valence electrons and free electrons.

Fatigue initiation is marked by the formation of PSB under cyclic loading, with the subsequent cyclic fatigue process being essentially the progressive evolution of these PSBs. Microcracks typically form only after approximately 90% of the total fatigue life has been consumed, followed by a rapid transition to fracture failure[3]. Therefore, it is evident that fatigue failure is an extended process during which interatomic bonding forces gradually weaken over time until they can no longer withstand the external loading forces[2,11].

Under static conditions, the relative positions and motion states of atoms remain relatively stable, with the electron cloud in a comparatively stable state. This stability allows us to employ techniques such as Transmission Electron Microscopy[12], Atomic Force Microscopy (AFM)[13], and X-ray diffraction[2] to deduce the arrangement and positioning of atoms, and subsequently calculate interatomic bonding forces using model theory.

However, quantifying the bonding forces of atoms in motion presents a significant challenge at present.

In the context of cyclic loading, the relative positions and motion states of atoms undergo periodic changes, leading to dynamic alterations in the distribution and interaction of electron clouds. These changes complicate the accurate measurement of the instantaneous values of interatomic bonding forces. Furthermore, the Heisenberg uncertainty principle of quantum mechanics dictates that it is impossible to simultaneously and precisely measure both the position and momentum of a microscopic particle, such as an electron, which also constrains the precise determination of interatomic bonding forces[14].

Additionally, cyclic loading can induce various defects and phase transitions within the material[15], which alter the relative positions and interactions between atoms, leading to changes in bonding forces. This further increases the complexity of measuring interatomic bonding forces.

**Quantum symbiotic relationship between magnetic interaction force and interatomic bonding force**

Ferromagnetic metals (mainly including iron, cobalt, nickel, and their alloys) are the most widely used metals and the materials that experience the most fatigue fractures. Their magnetism mainly originates from the electron spin magnetic moment[16]. Electron spin is an intrinsic angular momentum that couples with orbital motion, forming a spin orbit coupling effect that leads to degeneracy of electronic energy levels and discretization of magnetic energy levels, thereby affecting material magnetic properties[17] and the origin of interatomic magnetic interactions[18].

Given that the interatomic bonding force and magnetic interaction force are both related to the electronic motion state and are reflected in the same electron cloud configuration[19]. The change in atomic spacing between different layers causes a change in the electronic motion state. When the atomic spacing change is small, we can derive from the iron atom wave function that the average bonding force and magnetic interaction force between the atoms in the material layers are approximately linearly correlated with the average atomic spacing change: $\Delta F_C = K_C \Delta \bar{R}$ (Fig. 3e and Methods) and $\Delta F_m = K_m \Delta \bar{R}$ (Fig. 3d and Methods). From this, it can be inferred that there exists a quantum symbiotic relationship: $\Delta F_C = K \Delta F_m$ (Fig. 3c and Methods) between the interatomic bonding force and the magnetic interaction force. They interweave with each other, jointly shaping the atomic arrangement, electronic motion states, and macroscopic properties of materials.

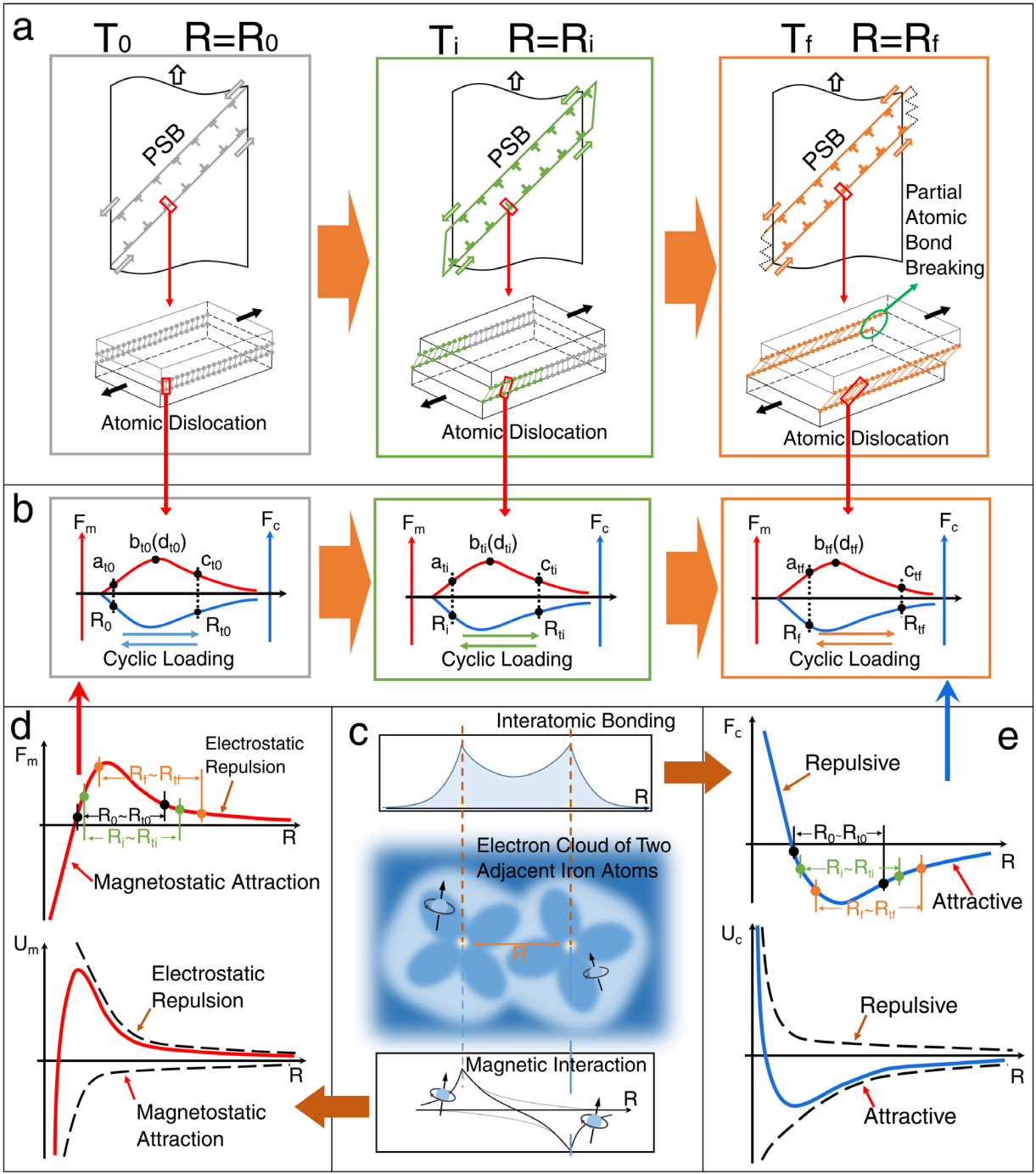

**Fig. 3 Schematic of the microscopic development process of fatigue and the synchronous relationship between interatomic bonding forces and magnetic forces. a**, PSB and interatomic dislocations of fatigue section. **b**, The relationship between interatomic bonding force and magnetic force varies with the change in atomic spacing. $R_0$, $R_{t0}$; $R_i$, $R_{ti}$; $R_f$ and $R_{tf}$ represent the atomic spacings at time points $T_0$, $T_i$, and $T_f$, respectively, under cyclic loading. The magnetic interactions at points 'a' and 'c' represent the magnetic interactions at the beginning of loading and at the maximum loading stress, respectively. Points 'b' and 'd' are the

peak points of magnetic interactions that occur during the loading process. As the loading time increases, atomic dislocations gradually increase, atomic spacing changes, and the magnetic interaction points 'a' and 'c' gradually shift. Peak change points 'b' and 'd', however, remain stationary. The bonding force $F_c$ and magnetic interaction force $F_m$ exhibit an increasing and then decreasing trend. The breakage of atomic bonds in the cross-section continues to increase, and fracture occurs when the overall bonding force of the cross-section becomes less than the applied force. **c**, Electron clouds, interatomic bonding, and magnetic interactions between adjacent iron atoms. **d**, The relationship between magnetic potential energy, as well as the magnetic interaction of electrons, and the atomic spacing of iron atoms. **e**, The relationship between bonding energy, as well as the bonding force, and the atomic spacing of iron atoms.

## Amplification of small fatigue signals via quantum spin correlation chain conduction mechanism

Under fatigue, the alterations in metals are predominantly confined to locally PSBs, which are induced by dislocations. These PSBs manifest as thin layers, and their cumulative total width within a region is relatively minor compared to the entire material, constituting only about 9.2% of the metal[20,21]. During the extended fatigue process, the changes attributable to each loading cycle may merely increment dislocations or atomic bond ruptures within a few atoms in a specific PSB layer[22]. The impact of these interatomic bonding force modifications on the overall stress and strain changes is also minimal. Capturing these localized microscopic changes is challenging with conventional measurement techniques, which is a primary reason for the difficulty in characterizing and analyzing fatigue.

It is widely acknowledged that metallic materials consist of atoms arranged in chain or lattice structures. The macroscopic magnetic properties of these materials are determined by the distribution of electron clouds and the spin states of electrons among these atoms[16]. Quantum spin correlation have been extensively utilized in the field of quantum information processing and quantum computing, primarily through electric field excitation to facilitate quantum state transfer and logic gate operations[23].

We proposed a magnetically excited quantum spin correlation conduction amplification for weak changes of fatigue process. Furthermore, magnetically excited quantum spin correlation can accomplish target spin state transfer solely through the known Heisenberg exchange interaction in quantum magnetism[24-26]. This theory achieves significant amplification of magnetic signals caused by fatigue through the quantum amplification effect. Specifically, during fatigue develops gradually, atoms within a quantum spin correlation can be perturbed due to dislocations or PSB, leading to minor changes in their electron clouds and spin states[27]. Through electron interactions, including interatomic bonding forces and exchange interactions, these changes are transmitted between adjacent atoms[28]. When an external magnetic field is applied to ferromagnetic metal materials, they accumulate and amplify along the magnetic field vector, triggering a chain reaction (Fig. 4d). This amplification phenomenon arises from the synergistic effects of quantum mechanical exchange interactions and external magnetic fields. Consequently, the subtle changes in the bonding forces and the magnetic interaction forces between a few atoms due to material fatigue are greatly amplified through the accumulation effect of quantum spin correlation, with the amplification factor precisely corresponding to the number of atomic layers involved in quantum spin correlation (Methods).

In summary, under external cyclic loading, the distribution of shared electron clouds within the material changes, leading to concurrent alterations in atomic bonding forces and magnetic interaction forces with symbiotic relationships. This further results in the reorientation of spin magnetic moments, macroscopically manifesting as changes in the material's magnetic properties. We harness the accumulation amplification effect of chain conduction to enhance the sensitivity of fatigue (interatomic bonding) observations. Upon the application of a magnetic field to ferromagnetic metals, after the magnetic signal variations undergo chain accumulation amplification and reach the material surface, the resulting magnetization of the material will induce a sequential chain conduction mechanism (Fig. 4d). In this process, magnetic induction flows from one segment to the next, with subtle alterations in the weakest link undergoing cumulative amplification, ultimately becoming strikingly evident in the final stage. Its amplification factor corresponds precisely to the number of PSB layers involved in the quantum spin correlation (Methods). By measuring the magnetic strength of materials, the progression of interatomic bonding forces and fatigue can be observed throughout the entire process.

**Significant enhancement in sensitivity**

In terms of specific implementation, probability of observing fatigue-induced electronic state (target states) changes has been enhanced by quantum spin correlation transferring layer by layer. The Observing Probability Amplitude (OPA) for changes in the electron cloud between a few atoms in the initial layer is relatively low. However, as the propagation through layers deepens, this OPA increases in a specific pattern (Fig. 4d). Employing the quantum amplitude amplification method[29], we analogize the observation probability amplitude of the first layer to the sine of an angle. With each subsequent layer, this angle increases, leading to a multi-level increase in the OPA of the target state. Ultimately, this transfer results in a significant amplification, thereby greatly enhancing the OPA and allowing us to efficiently monitor changes in interatomic bonding forces. The process of layer-by-layer transmission and its amplification effect, namely, the multiplicative increase in the OPA, was demonstrated in the fatigue tests of five ferromagnetic alloy materials conducted in this study. The displacement change of a two-electron spin system (a single metal bond) was amplified hundreds of times through the entire quantum spin correlation conduction chain (Methods), which significantly surpasses current microscopic image observation and S-N curve measurements. The actual observed magnetic strength changes in our experiments were, on median, 69.22 times (Fig. 4c) that of traditional mechanical quantities. For the first time, we have made a clear macroscopic observation of the dynamic changes in the magnetic effect (bonding force) throughout the entire (long) fatigue development process and noted a significant acceleration change prior to fracture.

**Macroscopic characterization of interatomic bonding forces: peak-to-peak magnetic intensity (termed as MagDrift)**

Concurrent with the external force altering atomic spacing, it also influences the degree of overlap and motion state of the electron clouds between atoms, resulting in alterations to the bonding force or magnetic interaction force. Within the elastic limit of the microscopic crystal structure, upon releasing the tensile force, atoms can revert to their pre-stretched positions[30]. Concurrently, the interatomic bonding force, magnetic interaction force, and surface magnetic strength also return to their initial states. However, once the elastic limit is surpassed, some atoms undergo plastic deformation, such as dislocations due to defects, preventing them from recovering to their original state[31]. In such cases, the interatomic forces, magnetic interactions, and surface magnetic strength cannot be restored.

During the fatigue, some atoms experience dislocations or bond ruptures in each cycle, precluding their return to original positions. These factors collectively result in the magnetic strength at the end of a cycle differing from the initial state, we call this change MagDrift. Consequently, the fatigue effect induces changes in MagDrift, with the MagDrift within each cycle reflecting the fatigue-induced alterations in interatomic bonding or magnetic interaction forces.

Through fatigue testing on five types of ferromagnetic materials, we discovered that the average rate of change of MagDrift ($k_{av}$) for the same material is highly linearly correlated with the number of loading cycles (Fig. 4 and Methods). The longer the fatigue life, the smaller the average rate of change, thereby confirming that the average rate of change of MagDrift ($k_{av}$) serves as a macroscopic characterization of the changes in interatomic bonding forces.

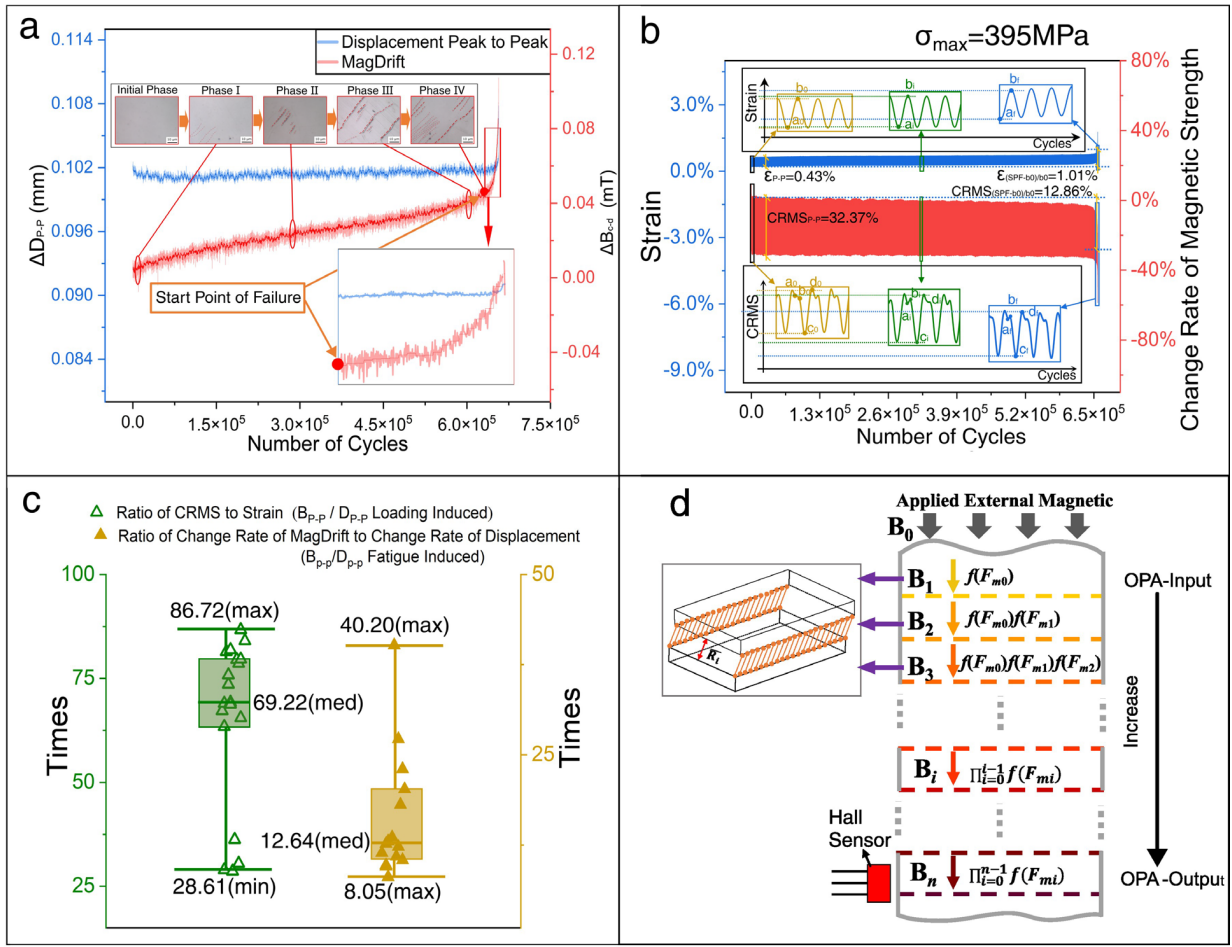

**Fig. 4 Schematic diagram of magnetic intensity variation trend and accumulation amplification effect during loading process based on quantum spin correlation conduction amplification mechanism. a**, Strain and change rate of magnetic strength under typical fatigue stress loads. **b**, Displacement amplitude and MagDrift changes under typical fatigue stress loads. **c**, Comparison of changes in magnetic strength and strain, MagDrift, and displacement amplitude under cyclic loading. The median rate of change in magnetic strength is 69.22 times higher than the median rate of change in strain, whereas the median change from beginning to start point of failure in MagDrift is 12.64 times greater than that of displacement. **d**, Schematic diagram of quantum spin conduction accumulation amplification.

## COOP as a determinant of bonding strength

In solid materials, the formation of metallic bonds depends on the constructive overlap of adjacent atomic electron clouds in the direction of maximum density to form shared electron pairs[32,33]. Consequently, the greater the depth of electron cloud overlap and the more extensive the atomic orbital overlap, the stronger the bonding strength becomes, making it more challenging to alter the interatomic bonding force. The degree of electron cloud overlap between two atoms is typically quantified by "Crystal Orbital Overlap Population" (COOP)[34]. The COOP is a measure of bond strength. Specifically, a higher COOP value indicates a greater number of bonding electrons between two atoms, which results in a stronger bond[35].

We observed that for materials predominantly composed of the same metal, such as iron and its alloys, the fitting slope $k_C$ (with a fitting degree greater than 0.9, p<0.0001) of the average change rate of interatomic bonding force is consistent, with a value of -1.54 (Fig. 5). Moreover, the $k_C$ for nickel is significantly lower than that for iron, and under the same fatigue load, the fatigue life of nickel is substantially longer than that of iron (Extended Data Fig. 2). This confirms that the difficulty of changing the interatomic bonding force is contingent upon the material's strength, which is primarily determined by the depth of electron cloud overlap of the metal atoms. The deeper the overlap, the more resistant the interatomic bonding force is to change, and the longer the fatigue life. Thus, $k_C$ serves as a macroscopic statistical representation of COOP and bond strength.

**The impact of ferromagnetic element content on macroscopic magnetic intensity**

The crystal structure of ferromagnetic alloy steels, such as Body-Centered Cubic (BCC) or Face-Centered Cubic (FCC) lattices, imparts a specific arrangement to magnetic atoms, which significantly influences the magnetic properties of the material[36]. Within these lattices, the atomic spacing and positional arrangement of ferromagnetic elements dictate the strength of exchange interactions, foundational to the emergence of ferromagnetism[37]. These interactions facilitate the alignment of adjacent atomic magnetic moments in the same direction and in parallel. Upon analyzing the elemental composition of ferromagnetic materials using an electron probe (Extended Data Fig. 5-9), we observed that the intercept constant $C$ of the fitting curve relating the average rate of change of MagDrift to the number of cycles was strongly correlated with the content of ferromagnetic elements, with higher content leading to a larger $C$ (Fig. 5b). The correlation is intuitive because an increase in the content of ferromagnetic elements results in higher magnetic permeability. This leads to a more pronounced change in magnetic strength under the same loading conditions and a greater magnitude of MagDrift from the same fatigue-induced changes.

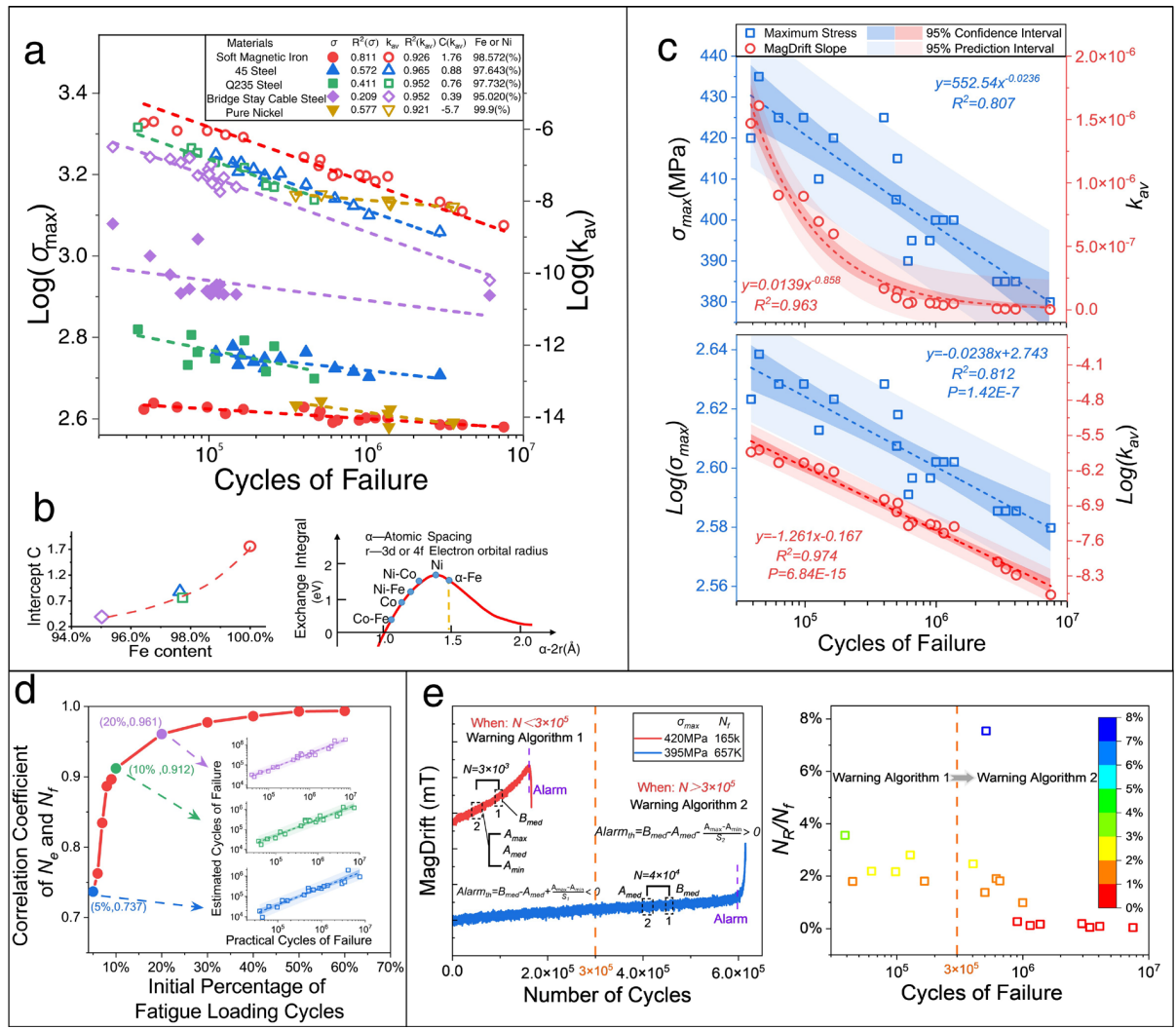

**Fig. 5 Analysis chart of experimental results. a**, Comparison of LogM-N and LogS-N curves for four ferromagnetic iron alloys. Use the average slope of the M-N fitting line of four ferromagnetic metals, -1.54, as the new slope ($k_c$). The LogM-N fitting degrees of them are all above 0.9 (p<0.0001), far greater than the LogS-N. **b**, Relationship between intercept $C$ of linear fitting line of LogM-N curve and iron content. The higher the iron content, the greater the intercept. **c**, Comparison of M-N and S-N curves for Soft Magnetic Iron specimens and comparison of LogM-N and LogS-N for Soft Magnetic Iron specimens. The M-N curve has a much higher fitting degree compared to the S-N curve. **d**, The estimated lifespan $N_e$ of Soft Magnetic Iron under different early cyclic loading conditions correlated with the actual lifespan $N_f$. The correlation coefficient between the estimated lifespan and the actual lifespan under 9% loading has already reached 0.907 (p<0.0001), which is significantly higher than the correlation coefficient between the maximum stress $\sigma_{max}$ and the actual lifespan $N_f$, which is -0.73. **e**, Early warning principle and effectiveness based on MagDrift. $S_1$ and $S_2$ are constants obtained through experimentation; for Soft Magnetic Iron, the value of $S_1$ and $S_2$ are 4.8 and 1.2 respectively. $A_{med}$ is the median of window 2, $A_{max}$ is the maximum value of window 2, $A_{min}$ is the minimum value of window 2, and $B_{med}$ is the median of window 1. $N_R$ is the remaining number of cycles to fatigue fracture.

**A deterministic model for fatigue life prediction**

Uncertainty factors, such as material doping[38], defects[39], and load variations[40], significantly influence the fatigue mechanical properties of materials, leading to considerable scatter in fatigue life prediction models that rely on microstructural and mechanical performance observations. We have developed an accurate correlation model between fatigue life and three core basic magneto physical parameters, leveraging the strong linear correlation between fatigue life ($N$), the average rate of change of interatomic bonding force ($k_{av}$), $k_C$ (macroscopic statistical representation of COOP), and the content of ferromagnetic elements ($C$).

$$\log N = \frac{1}{k_c}(\log|k_{av}| + C)$$

This deterministic fatigue life model incorporates three factors that were previously sources of high variability, such as the rate of change in interatomic bonding forces due to material doping, defects, load discrepancies, and bond overlap population. By accounting for these factors, the model mitigates the dispersion that arises from neglecting them during the construction of the model. Furthermore, in deriving the quantum symbiotic relationship between atomic bonding forces and magnetic interaction forces, a more precise depiction of atomic dynamic behavior is achieved by incorporating nonlinear effects through the application of nonlinear Schrödinger equations (Methods). This approach addresses the variability induced by uncertain factors.

Based on this formula, we propose an algorithm capable of accurately assessing the ultimate fatigue life of materials using early data. For the first time, the prediction accuracy of the original fatigue life has been enhanced to a fitting degree of 0.9 ($p<0.0001$) or above (Fig. 5d), significantly surpassing the linear fitting degree of the S-N curve.

**Fatigue failure warning**

The limitations in sensitivity and precision have constrained the accuracy of existing fatigue measurement methods, leading to premature or delayed alarms. Consequently, a reliable fatigue failure warning technology has not yet been fully realized. HR-DIC and TEM microscopic measurements reveal that the early PSB slip marks during the fatigue process almost entirely coincide with the final fracture position. There is a significant correlation between material mechanical properties, such as yield strength, ultimate tensile strength, and fatigue strength, and the early local slip of PSB[3,4]. Thus, early PSB changes can be utilized for fatigue life assessment.

Leveraging the high sensitivity of quantum spin correlation enhancement to fatigue changes, we can accurately capture key signals from materials during fatigue accumulation and ultimately achieve 100% fatigue fracture warning for 193 specimens of five different ferromagnetic materials across 3,700 testing hours. MagDrift undergoes significant changes a considerable time before material fracture occurs. Leveraging this characteristic, we propose a two-level warning system for material fatigue failure. The first level is a preliminary warning issued when only 10% of the material fatigue life remains, indicating that the material can still function normally, but it is about to enter a dangerous state. The second level is an extreme warning triggered when the remaining fatigue life is between 1.8% and 7.54%. Through this system, all test specimens successfully emitted warning signals prior to fatigue fracture (Fig. 5e). This study has, for the first time, achieved effective early warning of material fatigue fractures (Supplementary Video 1).

**Discussion**

The quantum spin correlation conduction amplification technology proposed in this study demonstrates transformative potential in ferromagnetic metal fatigue assessment. By leveraging magnetic field-induced quantum spin correlation conduction, we have achieved macroscopic detection of subtle quantum symbiotic interactions between interatomic bonding forces and magnetic interactions. The core innovation lies in the quantum spin accumulation amplification effect, which enables cumulative amplification of localized microscopic changes into detectable macroscopic magnetic property variations. This spin correlation conduction mechanism not only enhances detection sensitivity significantly (Fig. 4c), but also establishes a foundation for precise fatigue life prediction.

Our experimental results reveal three critical advances. First, quantum spin correlation exhibit exceptional responsiveness to minute bonding force alterations, with atomic bond displacements being amplified by up to 86.72 times through multi-layer conduction (Fig. 4c-d). This breakthrough enables the first macroscopic observation of dynamic magnetic effects (bonding force evolution) across full fatigue progression. Second, the developed model integrating electron-cloud-overlap constant of materials, ferromagnetic content, and bonding force degradation rates achieves unprecedented prediction accuracy ($R^2>0.9$, $p<0.0001$), resolving the century-old dispersion challenge in S-N curve methodologies. Finally, the two-tier warning system demonstrated 100% pre-fracture alert accuracy across 193 specimens, providing a critical safety buffer of 1.8-7.54% remaining life for emergency response.

Notably, the strong correlation between macroscopic statistical of COOP values and material fatigue resistance provides new insights into fatigue mechanism interpretation. Materials with deeper electron cloud overlap pure nickel exhibited longer fatigue lives compared to those with pure iron. Furthermore, the ferromagnetic content-dependent amplification constant $C$ (Fig. 5b) offers a quantitative framework for material selection in critical applications.

While current validation focuses on ferromagnetic systems, the underlying quantum spin correlation accumulation amplification principle shows promise for broader applications. Future directions include integration with machine learning for real-time remaining life estimation and extension to non-ferromagnetic systems through spin-orbit coupling modulation.

This work bridges the critical gap between quantum-scale phenomena and macroscopic engineering safety, establishing a new paradigm for predictive maintenance in aerospace, civil infrastructure, and energy systems.

**Publisher's note** Springer Nature remains neutral with regard to jurisdictional claims in published maps and institutional affiliations.

# Methods

**Preparation of test specimens and experimental procedures**

The materials employed in the experiment included Soft Magnetic Iron (DT4C), Q235 Steel (PB240), 45 Steel (C45E4), Pure Nickel (Nickel 201), and Bridge Stay Cable Steel (BSCS). DT4C, PB240, C45E4, and Nickel 201 were all 16mm-diameter round rods, while the Bridge Stay Cable Steel was 7 mm in diameter. Before loading, the central sections of the specimens were machined into an arc form utilizing Computer Numerical Control (CNC) techniques. The narrowest section of the Soft Magnetic Iron (DT4C), Q235 Steel (PB240), 45 Steel (C45E4), and Pure Nickel (Nickel 201) samples measured 8mm, while the Bridge Stay Cable Steel's sample had a narrowest section of 4mm (Extended Data Table 1).

MTS Landmark Model 370.10 electro-hydraulic servo mechanical test system; five N35 magnets; the Hall sensor adopts EQ-730L, which is composed of an InAs Quantum Well Hall Element and a signal processing IC chip in a package. Supply voltage from 3.0V to 5.5V at single power supply (Extended Data Table 1). The data was collected using the DH5922N dynamic signal analysis and acquisition system from Donghua Test.

We designed a material fatigue observation test device based on quantum spin correlation (Extended Data Fig. 1a), and conducted static load and fatigue tests based on quantum spin correlation monitoring using Soft Magnetic Iron (DT4C), Q235 Steel (PB240), 45 Steel (C45E4), Pure Nickel (Nickel 201), and Bridge Stay Cable Steel. All fatigue tests have a loading frequency of 20Hz and a stress ratio of 0. During this period, the magnet and Hall sensor were kept out of contact with the specimen, and the distance between them remained constant.

Both ends of the specimen were clamped on the high-frequency fatigue test machine of MTS Landmark Model 370.10 electro-hydraulic servo mechanical test system, and five N35 magnets were vertically superimposed and adsorbed to the upper fixture of the universal testing machine.

The Hall sensor is positioned in parallel, approximately 1 cm below the arc-shaped section of the specimen, maintaining a vertical clearance of approximately 2 mm from the specimen. The distance between the magnet and the Hall sensor remains constant, specifically when the universal testing machine is in operation, with the upper fixture stationary and the lower fixture subject to loading, as illustrated in Extended Data Fig. 1a.

Fatigue test parameters are set: loading frequency is 20 Hz, data acquisition frequency is 128 Hz, cyclic loading mode is adopted by force control, and the minimum stress is 0 MPa. When the specimen fatigue breaks, the data collection is stopped and the data is processed.

**Detailed statistical method for fatigue accidents and fatalities**

In an effort to provide a comprehensive tally of accidents and fatalities stemming from metal fatigue, we utilized targeted keywords—"List of bridge failures", "List of elevator accidents", "List of industrial disasters", "List of railway accidents", "Aviation safety network", "List of marine disasters in the 21st century", "List of marine disasters in the 20th century" and "List of building and structure collapses"—to extract data from "Wikipedia" on incidents across various sectors, including bridges, elevators, industry, railways, aviation, marine, and construction. We then compiled a statistical database documenting these accidents over the past century, which includes the time, location, number of fatalities, and detailed descriptions of each event. To safeguard the integrity of our data, we meticulously reviewed adjacent records to identify and remove any duplicates.

Subsequently, we utilized a series of keywords to compare and filter the causes of accidents across various industries. For the bridge industry, we used keywords "metal fatigue", "fatigue cracks", "metal corrosion and fatigue", "structural performance deterioration", "structural collapse", "overload", "suspension cable fracture" and "steel frame fracture". In the elevator industry, we screened for causes using keywords "cable breakage", "steel cable breakage", "overload" and "mechanical failure". For the industrial sector, we applied keywords "ultimate stress", "corrosion cracking", "mechanical failure", "train derailment", "cable wear" and "cable fracture". In the railway industry, we searched for causes with keywords "train derailment", "axle failure", "engine failure", "rail breakage", "boiler failure", "brake wear" and "overload". For the ship industry, we compared and screened causes using keywords "overload", "outdated ship", "engine failure", "propeller shaft fracture", "boiler failure", "deck rupture" and "ship vibration". In the construction industry, we use keywords "bridges", "masts", "aircraft racks", "steel structures", "cranes" and "stadium roofs" to screen the structural types. Finally, for the aviation industry, we used keywords "wing flutter", "engine failure", "overload", "mechanical failure", "landing gear failure", "fuselage bending" and "engine vibration" to compare and screen the causes of accidents. For the selected accidents, we manually searched each accident on "Google" and "Wikipedia", analyzed the usage time of the buildings or mechanical equipment before the accident occurred, and examined photos of the accident scenes. Then, we used keywords "metal fatigue", "fatigue fracture", "fatigue crack", "fatigue cracking" and "fatigue failure" to rescreen the accidents, ultimately classifying those that matched these keywords as metal fatigue-related accidents.

In addition, we utilized keywords "metal fatigue", "fatigue fracture", "fatigue crack", "fatigue cracking" and "fatigue failure" to compare and screen the causes of aviation accidents, extracting accident records that matched these keywords and directly classifying them as accidents caused by metal fatigue. Subsequently, we manually sorted and integrated the selected accidents in chronological order, ultimately obtaining a compilation of accidents caused by metal fatigue in the bridge, elevator, industrial, railway, aviation, ship and construction industries spanning from 1900 to 2024.

Finally, we classified the accidents caused by metal fatigue and extracted the number of deaths from each accident's raw data, and then calculated the cumulative number of deaths caused by metal fatigue in each industry every five years to obtain the total number of deaths caused by metal fatigue in each industry over the past century, as well as the number of deaths occurring every five years.

**Calculation method for maintenance costs of various industries in 2023**

Infrastructure like bridges, elevators, railways, ships, and airplanes is constantly affected by metal fatigue due to multiple factors, including long-term heavy loads, vibrations, and corrosion, leading to a gradual decline in their performance. Therefore, metal fatigue, as the main driver of maintenance costs for these infrastructures, cannot be ignored in terms of its impact. In order to accurately and comprehensively reflect the latest trends in maintenance costs within the industry, especially the impact of metal fatigue on these costs, we have adopted a systematic methodological approach. This involved conducting detailed multi-database surveys and utilizing predefined terminology "Global 'Bridge', 'Lift', 'Railway', 'Ship', and 'Aviation' maintenance market size in 2023" to conduct a detailed statistical analysis on the maintenance cost data related to metal fatigue in the bridge, elevator, railway, ship, and aircraft industries for the year 2023.

After rigorous screening, we found that metal fatigue significantly impacts maintenance costs across various industries. The maintenance cost for bridges due to metal fatigue is as high as $5.4 billion, while the elevator industry costs approximately $0.406 billion. The maintenance costs for the railway and shipping industries reached $4.328 billion and $24.202 billion, respectively, and the aircraft industry had the highest maintenance costs, reaching $44.566 billion. These data not only reveal the direct impact of metal fatigue on maintenance costs but also reflect the economic pressures that different industries face in addressing metal fatigue issues.

**Method for surveying the popularity of fatigue research literature**

Using a systematic methodology, literature related to the keywords "Fatigue", "PSB", "Accumulation of dislocations", "S-N Curve" and "Evolution of temperature entropy" was retrieved from the "Web of Science" and "Engineering Village" databases to comprehensively evaluate the popularity of fatigue research literature.

**Theoretical model of the relationship between interlayer bonding force, magnetic force, and atomic spacing variation under cyclic loading**

It is generally believed that bonding energy manifests as Coulomb repulsion energy and attraction energy. When the distance between atoms is close, the repulsive force between negatively charged electrons dominates, with the predominant energy being manifested as repulsive energy. When the atomic distance is far, the main energy is manifested as attractive energy. In a free state, energy is at an equilibrium point where the sum of repulsive and attractive energies is minimized.

In the study, only the tensile state of the material, commencing from its free state, is considered, and specifically, only the scenario where the distance between atoms undergoes relatively minor changes is taken into account.

From the measurement and simulation images of iron atom electron clouds, it can be seen that electrons are distributed around the atom in the form of probability density functions of wavefunctions. According to the representation method in solid-state physics (Schrödinger equation), we express the bonding energy in the form of a wavefunction.

$$U_C = \frac{e^2}{R} + \iint \psi_{ab}^* \cdot e^2 \left(\frac{1}{r_{12}} - \frac{1}{r_{b1}} - \frac{1}{r_{a2}}\right) \cdot \psi_{ab} d\tau_1 d\tau_2 \quad (1)$$

Where $R$ is the distance between atoms, $r_{12}$ is the distance between electrons, $r_{b1}$ and $r_{a2}$ are the distances between atoms and adjacent electrons. $\psi_{ab}$ and $\psi_{ab}^*$ are the wave function and conjugate of the electrons that form the bonding energy, $e$ representing the charge distribution between the electrons that form the bonding energy and the atom.

At present, there are various models for the bonding energy between iron atoms, such as the localized electron model, the itinerant electron model, the band model, etc. However, the basic forms of these models can all be represented by the above formula, with only minor differences.

If it is the bonding energy between multiple pairs of atoms, we will write Equation 1 as follows:

$$U_C = \Sigma \left( \frac{e^2}{R_{ij}} + \iint \psi_{ab}^* \cdot e^2 \left( \frac{1}{r_{ij}} - \frac{1}{r_{bi}} - \frac{1}{r_{aj}} \right) \cdot \psi_{ab} d\tau_1 d\tau_2 \right) \qquad (2)$$

Where $R_{ij}$ is the distance between atoms $i$ and $j$, $r_{ij}$ is the distance between electrons $i$ and $j$, $r_{bi}$ and $r_{aj}$ are the distance between atoms $a$ and adjacent electron $i$ and atoms $b$ and adjacent electron $j$, respectively.

Note that the bonding energy of a material increases monotonically from the lowest energy state when it is stretched. We use a function of the average atomic spacing between the two layers of the material's tensile cross-section, denoted as $\bar{r}_c$, to simplify the equivalent overall wave function.

$$\frac{1}{\bar{r}_c} \equiv \Sigma \left( \frac{1}{R_{ij}} + \iint \psi_{ab}^* \cdot \left( \frac{1}{r_{ij}} - \frac{1}{r_{bi}} - \frac{1}{r_{aj}} \right) \cdot \psi_{ab} d\tau_1 d\tau_2 \right) \qquad (3)$$

Where $\bar{r}_c$ actually represents the average Coulomb interaction distance between two layers of atoms

By substituting equation 3 into equation 2, we obtain the equation 4:

$$U_C = \frac{e^2}{\bar{r}_c} \qquad (4)$$

Taking the derivative of $U_C$ with respect to average interatomic distance $\bar{R}$ of two layers of atoms yields:

$$F_C = \frac{dU_C}{d\bar{R}} = -\frac{e^2}{\bar{r}_c^2} \frac{d\bar{r}_c}{d\bar{R}} \qquad (5)$$

Where $F_C$ represents the average bonding force between two layers of atoms.

By expanding $\bar{r}_c$ in terms of $\bar{R}$ using Taylor series expansion, we can obtain: $\bar{r}_c = C_{C0} + C_{C1}\bar{R} + C_{C2}\bar{R}^2 + \cdots$, Where $C_{C0}$, $C_{C1}$ and $C_{C2}$ are constants.

Ignoring higher-order terms and only considering the case where the atomic spacing increases slightly from its lowest energy position, where $\bar{r}_c$ linearly changes with respect to $\bar{R}$, we can obtain:

$$F_C = -\frac{e^2 C_{C1}}{(C_{C0} + C_{C1}\bar{R})^2} = -\frac{d_{C0}}{(\bar{R} + d_{C1})^2} \qquad (6)$$

Where $d_{C1} = \frac{C_{C0}}{C_{C1}}$ and $d_{C0} = \frac{e^2}{C_{C1}}$ are constants.

So, we can use a function of atomic spacing to express the change in average interatomic bonding force during tensile fatigue.

Magnetic interactions are relatively more complex, but here they are simply calculated using the same formalism as that for the bonding force. The magnetic exchange energy, which arises from the exchange interaction between electrons of two atoms as well as between electrons and atomic nuclei, is expressed by a wave function.

$$U_m = \iint \varphi_{ab}^* \cdot m \left( \frac{1}{r_{12}} - \frac{1}{r_{b1}} - \frac{1}{r_{a2}} \right) \cdot \varphi_{ab} d\tau_1 d\tau_2 \qquad (7)$$

The wave functions of the electrons that contribute to magnetic exchange energy are represented by $\varphi_{ab}$ and their conjugates by $\varphi_{ab}^*$, respectively. $m$ represents the combined charge and magnetic moment of the electrons and atoms that form magnetic exchange energy.

For the magnetic exchange energy between multiple pairs of atoms, we will write equation 7 as:

$$U_m = \sum \iint \varphi_{ab}^* \cdot m \left( \frac{1}{r_{ij}} - \frac{1}{r_{bi}} - \frac{1}{r_{aj}} \right) \cdot \varphi_{ab} d\tau_1 d\tau_2 \tag{8}$$

Where $r_{ij}$ is the distance between electrons $i$ and $j$, $r_{bi}$ and $r_{aj}$ are the distance between atoms $a$ and adjacent electron $i$ and atoms $b$ and adjacent electron $j$, respectively.

For a two-layer atomic structure, we obtain:

$$\frac{1}{\bar{r}_m} \equiv \sum \iint \varphi_{ab}^* \cdot \left( \frac{1}{r_{ij}} - \frac{1}{r_{bi}} - \frac{1}{r_{aj}} \right) \cdot \varphi_{ab} d\tau_1 d\tau_2 \tag{9}$$

By substituting equation 9 into equation 8, we obtain:

$$U_m = \frac{m}{\bar{r}_m} \tag{10}$$

The overline notation $\bar{r}_m$ actually represents the equivalent magnetic interaction distance between two layers of atoms. Similarly, $\bar{r}_m$ is a function of the average distance $\bar{R}$ between these two layers of atoms. Under other constant conditions, once the average distance $\bar{R}$ between the layers is determined, $\bar{r}_m$ takes on a unique value.

By taking the derivative of $U_m$ on $\bar{R}$, we can obtain:

$$F_m = \frac{dU_m}{d\bar{R}} = -\frac{m}{\bar{r}_m^2} \frac{d\bar{r}_m}{d\bar{R}} \tag{11}$$

Where $F_m$ represents the average magnetic force between two layers of atoms.

By expanding $\bar{r}_m$ in terms of $\bar{R}$ using Taylor series expansion, we can obtain: $\bar{r}_m = C_{m0} + C_{m1}\bar{R} + C_{m2}\bar{R}^2 + \cdots$, Where $C_{m0}$, $C_{m1}$ and $C_{m2}$ are constants.

Similarly, ignoring higher-order terms and only considering $\bar{r}_m$ linearly changes with respect to $\bar{R}$, we can obtain:

$$F_m = -\frac{mC_{m1}}{(C_{m0} + C_{m1}\bar{R})^2} = -\frac{d_{m0}}{(\bar{R} + d_{m1})^2} \tag{12}$$

Where $d_{m1} = \frac{C_{m0}}{C_{m1}}$ and $d_{m0} = \frac{m}{C_{m1}}$ are constants.

From this, we can consider that the average bonding force and magnetic exchange force between adjacent layers of atoms in ferromagnetic materials are homologous effects, and both exhibit an inverse square relationship with the average atomic spacing between layers. Therefore, in cases where $\bar{R}$ changes very slightly, an increase in $\bar{R}$ leads to a decrease in both the bonding force and the magnetic force (we are only considering the effect of an increase in $\bar{R}$).

When there is an external cyclic loading, let the force exerted by the loading on the two layers of atoms be denoted as $f(\sigma)$, and the interlayer magnetic force caused by the loading be denoted as $g(\sigma)$. The function form of $g(\sigma)$ is relatively complex, and since it can be omitted in the subsequent derivation, its specific form will not be considered here.

During cyclic loading, the stress gradually increases to the maximum level and then decreases back to zero, repeating the cycle around an intermediate stress value.

When external loading is $\sigma_m - \sigma_L$, the combined effects of bonding force and loading action are as follows:

$$F_{C0} - f(\sigma_m - \sigma_L) \tag{13}$$

The combined effects of magnetic force and loading action are as follows:

$$F_{m0} - g(\sigma_m - \sigma_L) \tag{14}$$

Where $F_{C0} = -\frac{d_{C0}}{(\bar{R}_0+d_{C1})^2}$, $F_{m0} = -\frac{d_{m0}}{(\bar{R}_0+d_{m1})^2}$.

When external loading is $\sigma_m + \sigma_L$, the combined effects of bonding force and loading action are as follows:

$$F_{C1} - f(\sigma_m + \sigma_L) \tag{15}$$

The combined effects of magnetic force and loading action are as follows:

$$F_{m1} - g(\sigma_m + \sigma_L) \tag{16}$$

Where $F_{C1} = -\frac{d_{C0}}{(\bar{R}_1+d_{C1})^2}$, $F_{m1} = -\frac{d_{m0}}{(\bar{R}_1+d_{m1})^2}$, $\sigma_m$ represents the average force during external cyclic loading, $\sigma_L$ is the fluctuation value of force within this cycling. $F_{C0}$ is the average bonding force inside the material when the external loading is $\sigma_m - \sigma_L$, specifically at the onset of fatigue loading, and $F_{C1}$ is the average bonding force inside the material when the initial loading is set to $\sigma_m + \sigma_L$, also at the onset of fatigue loading. Similarly, $F_{m0}$ is the average magnetic force inside the material when the external loading is $\sigma_m - \sigma_L$, specifically at the onset of fatigue loading, and $F_{m1}$ is the average magnetic force inside the material when the external loading is $\sigma_m + \sigma_L$, also at the onset of fatigue loading. Furthermore, $\bar{R}_0$ denotes the average atomic spacing inside the material when the external loading is $\sigma_m - \sigma_L$, specifically at the onset of fatigue loading, while $\bar{R}_1$ represents the average atomic spacing inside the material when the initial loading is $\sigma_m + \sigma_L$, also at the onset of fatigue loading.

Subtracting equation 13 from equation 15 results in the difference in bonding force, which represents the peak-to-peak value of the bonding force during loading, specifically at the onset of fatigue loading:

$$F_{CPP} = F_{C1} - f(\sigma_m + \sigma_L) - [F_{C0} - f(\sigma_m - \sigma_L)] = f(\sigma_m - \sigma_L) - f(\sigma_m + \sigma_L) - \frac{d_{C0}}{(\bar{R}_1+d_{C1})^2} + \frac{d_{C0}}{(\bar{R}_0+d_{C1})^2} \tag{17}$$

Subtracting equation 14 from equation 16 results in the difference in magnetic force, which represents the peak-to-peak value of the magnetic force during loading, specifically at the onset of fatigue loading:

$$F_{mPP} = F_{m1} - g(\sigma_m + \sigma_L) - [F_{m0} - g(\sigma_m - \sigma_L)] = g(\sigma_m - \sigma_L) - g(\sigma_m + \sigma_L) - \frac{d_{m0}}{(\bar{R}_1+d_{m1})^2} + \frac{d_{m0}}{(\bar{R}_0+d_{m1})^2} \tag{18}$$

Because the diameter in the middle of the specimen is smaller compared to the surrounding area, fatigue is most likely to occur at the middle position. We assume that fatigue dislocations arise between the two layers of atoms in the middle (although there may be more layers involved, it can be assumed that the two weakest layers of dislocations ultimately determine fracture).

When considering material fatigue, the average atomic spacing $\bar{R}_0$ increases by $\Delta\bar{R}$, where $\Delta\bar{R}$ represents the fatigue-induced dislocation change that occurs in the specimen under external cyclic loading, and it is assumed that $\bar{R}_0 \gg \Delta\bar{R}$.

For the external loading $\sigma_m - \sigma_L$ the combined effects of bonding force and loading action can be written as follows:

$$F'_{C0} - f(\sigma_m - \sigma_L) \tag{19}$$

And the combined effects of magnetic force and loading action are as follows:

$$F'_{m0} - g(\sigma_m - \sigma_L) \tag{20}$$

Where $F'_{C0} = -\frac{d_{C0}}{(\bar{R}_0+\Delta\bar{R}+d_{C1})^2}$, $F'_{m0} = -\frac{d_{m0}}{(\bar{R}_0+\Delta\bar{R}+d_{m1})^2}$.

When the external loading is $\sigma_m + \sigma_L$, and also considering the increase in the distance between the two layers of atoms from their initial equilibrium position $\bar{R}_1$ by $\Delta\bar{R}$, and $\bar{R}_1 \gg \Delta\bar{R}$.

Then the combined effects of bonding force and loading action can be written as follows:

$$F'_{C1} - f(\sigma_m + \sigma_L) \tag{21}$$

And the combined effects of magnetic force and loading action are as follows:

$$F'_{m1} - g(\sigma_m + \sigma_L) \tag{22}$$

Where $F'_{C1} = -\frac{d_{C0}}{(\bar{R}_1+\Delta\bar{R}+d_{C1})^2}$, $F'_{m1} = -\frac{d_{m0}}{(\bar{R}_1+\Delta\bar{R}+d_{m1})^2}$.

The difference obtained by subtracting 19 from 21 represents the actual change in bonding force that occurs due to fatigue, specifically when a fatigue dislocation takes place during loading:

$$F'_{CPP} = F'_{C1} - f(\sigma_m + \sigma_L) - [F'_{C0} - f(\sigma_m - \sigma_L)] = f(\sigma_m - \sigma_L) - f(\sigma_m + \sigma_L) - \frac{d_{C0}}{(\bar{R}_1+\Delta\bar{R}+d_{C1})^2} + \frac{d_{C0}}{(\bar{R}_0+\Delta\bar{R}+d_{C1})^2} \tag{23}$$

The difference obtained by subtracting 20 from 22 represents the actual change in magnetic force that occurs due to fatigue, specifically when a fatigue dislocation takes place during loading:

$$F'_{mPP} = F'_{m1} - g(\sigma_m + \sigma_L) - [F'_{m0} - g(\sigma_m - \sigma_L)] = g(\sigma_m - \sigma_L) - g(\sigma_m + \sigma_L) - \frac{d_{m0}}{(\bar{R}_1+\Delta\bar{R}+d_{m1})^2} + \frac{d_{m0}}{(\bar{R}_0+\Delta\bar{R}+d_{m1})^2} \tag{24}$$

Subtracting equation 23 from equation 17 yields the relationship between the change in fatigue dislocation $\Delta\bar{R}$ and the change in average bonding force during actual loading:

$$\Delta F_C = F'_{CPP} - F_{CPP} = \left[-\frac{d_{C0}}{(\bar{R}_1+\Delta\bar{R}+d_{C1})^2} + \frac{d_{C0}}{(\bar{R}_0+\Delta\bar{R}+d_{C1})^2}\right] - \left[-\frac{d_{C0}}{(\bar{R}_1+d_{C1})^2} + \frac{d_{C0}}{(\bar{R}_0+d_{C1})^2}\right] =$$

$$\left[\frac{d_{C0}}{(\bar{R}_1+d_{C1})^2} - \frac{d_{C0}}{(\bar{R}_1+\Delta\bar{R}+d_{C1})^2}\right] - \left[\frac{d_{C0}}{(\bar{R}_0+d_{C1})^2} - \frac{d_{C0}}{(\bar{R}_0+\Delta\bar{R}+d_{C1})^2}\right] = \frac{d_{C0}(\Delta\bar{R}^2+2\bar{R}_1\Delta\bar{R}+2\Delta\bar{R}d_{C1})}{(\bar{R}_1+d_{C1})^2(\bar{R}_1+\Delta\bar{R}+d_{C1})^2} -$$

$$\frac{d_{C0}(\Delta\bar{R}^2+2\bar{R}_0\Delta\bar{R}+2\Delta\bar{R}d_{C1})}{(\bar{R}_0+d_{C1})^2(\bar{R}_0+\Delta\bar{R}+d_{C1})^2} \tag{25}$$

Due to $\bar{R}_0, \bar{R}_1 \gg \Delta\bar{R}$, $\Delta\bar{R}^2$ in the above equation can be neglected. Therefore, equation 25 can be simplified as:

$$\Delta F_C \approx \frac{d_{C0}(2\bar{R}_1\Delta\bar{R}+2\Delta\bar{R}d_{C1})}{(\bar{R}_1+d_{C1})^2(\bar{R}_1+\Delta\bar{R}+d_{C1})^2} - \frac{d_{C0}(2\bar{R}_0\Delta\bar{R}+2\Delta\bar{R}d_{C1})}{(\bar{R}_0+d_{C1})^2(\bar{R}_0+\Delta\bar{R}+d_{C1})^2} = 2\Delta\bar{R}d_{C0}\left[\frac{1}{(\bar{R}_1+d_{C1})(\bar{R}_1+\Delta\bar{R}+d_{C1})^2} - \frac{1}{(\bar{R}_0+d_{C1})(\bar{R}_0+\Delta\bar{R}+d_{C1})^2}\right] \quad (26)$$

Simplify equation 26 to:

$$\Delta F_C = K_C \Delta\bar{R} \quad (27)$$

Where $K_C = 2d_{C0}\left[\frac{1}{(\bar{R}_1+d_{C1})(\bar{R}_1+\Delta\bar{R}+d_{C1})^2} - \frac{1}{(\bar{R}_0+d_{C1})(\bar{R}_0+\Delta\bar{R}+d_{C1})^2}\right]$, $\Delta F_C$ is the average variation of interatomic bonding force.

Due to $\bar{R}_1$ and $\bar{R}_0$ are related to the external force magnitude during loading and the inherent properties of the material, once the material and loading force are specified, the averages of $\bar{R}_1$ and $\bar{R}_0$ take on unique values. Given that $\bar{R}_1 - \bar{R}_0 \gg \Delta\bar{R}$, $\Delta\bar{R}$ has a negligible effect on the denominator term in $K_C$, thus allowing us to approximate $K_C$ as a constant. When the material undergoes fatigue dislocations, the change in average bonding force is solely dependent on the change in average atomic spacing.

Similarly, subtracting equation 24 from equation 18 yields the relationship between the change in fatigue dislocation $\Delta\bar{R}$ and the change in average magnetic force during actual loading:

$$\Delta F_m = F'_{mPP} - F_{mPP} = \left[-\frac{d_{m0}}{(\bar{R}_1+\Delta\bar{R}+d_{m1})^2} + \frac{d_{m0}}{(\bar{R}_0+\Delta\bar{R}+d_{m1})^2}\right] - \left[-\frac{d_{m0}}{(\bar{R}_1+d_{m1})^2} + \frac{d_{m0}}{(\bar{R}_0+d_{m1})^2}\right] = \left[\frac{d_{m0}}{(\bar{R}_1+d_{m1})^2} - \frac{d_{m0}}{(\bar{R}_1+\Delta\bar{R}+d_{m1})^2}\right] - \left[\frac{d_{m0}}{(\bar{R}_0+d_{m1})^2} - \frac{d_{m0}}{(\bar{R}_0+\Delta\bar{R}+d_{m1})^2}\right] = \frac{d_{m0}(\Delta\bar{R}^2+2\bar{R}_1\Delta\bar{R}+2\Delta\bar{R}d_{m1})}{(\bar{R}_1+d_{m1})^2(\bar{R}_1+\Delta\bar{R}+d_{m1})^2} - \frac{d_{m0}(\Delta\bar{R}^2+2\bar{R}_0\Delta\bar{R}+2\Delta\bar{R}d_{m1})}{(\bar{R}_0+d_{m1})^2(\bar{R}_0+\Delta\bar{R}+d_{m1})^2} \quad (28)$$

Simplify equation 28 to:

$$\Delta F_m \approx \frac{d_{m0}(2\bar{R}_1\Delta\bar{R}+2\Delta\bar{R}d_{m1})}{(\bar{R}_1+d_{m1})^2(\bar{R}_1+\Delta\bar{R}+d_{m1})^2} - \frac{d_{m0}(2\bar{R}_0\Delta\bar{R}+2\Delta\bar{R}d_{m1})}{(\bar{R}_0+d_{m1})^2(\bar{R}_0+\Delta\bar{R}+d_{m1})^2} \approx 2\Delta\bar{R}d_{m0}\left[\frac{1}{(\bar{R}_1+d_{m1})(\bar{R}_1+\Delta\bar{R}+d_{m1})^2} - \frac{1}{(\bar{R}_0+d_{m1})(\bar{R}_0+\Delta\bar{R}+d_{m1})^2}\right] \quad (29)$$

Simplify equation 29 to:

$$\Delta F_m = K_m \Delta\bar{R} \quad (30)$$

Where $K_m = 2d_{m0}\left[\frac{1}{(\bar{R}_1+d_{m1})(\bar{R}_1+\Delta\bar{R}+d_{m1})^2} - \frac{1}{(\bar{R}_0+d_{m1})(\bar{R}_0+\Delta\bar{R}+d_{m1})^2}\right]$, is also regarded as a constant. $\Delta F_m$ represents the average variation of magnetic force. Then, the change in average magnetic force when a material experiences fatigue dislocations is only related to the change in average atomic spacing too.

Combining equation 27 and equation 30, the relationship between bonding force and magnetic force can be obtained:

$$\Delta F_C = K \Delta F_m \quad (31)$$

Where $K = \frac{K_C}{K_m}$.

**Theoretical modeling of quantum spin correlation process and analysis of amplification mechanism**

When an external magnetic field is applied to ferromagnetic metals, the magnetic induction intensity is transmitted layer by layer along a certain direction due to the (virtual) interlayer magnetic interaction. Assuming that interlayer magnetic interaction is the sole determining factor for the transfer of magnetic field strength within the magnetic field, and neglecting the magnetic field strength transmitted through air, we can present the relationship between the induced magnetic field strength, the applied magnetic field strength, and the interlayer magnetic interaction force after the external magnetic field has passed through one layer (Fig. 4d):

$$B_1 = B_0 \cdot f(F_{m0}) \tag{32}$$

Where $B_0$ represents the external magnetic field strength, $B_1$ represents the induced magnetic field strength of the first layer, $F_{m0}$ denotes the interlayer magnetic interaction, and $f(F_{m0})$ is a continuous function of the interlayer magnetic interaction force.

The magnetic field strength of the subsequent layer is induced by the interlayer magnetic interaction force from the preceding layer, thus we have:

$$B_2 = B_1 \cdot f(F_{m1}) \tag{33}$$

Where $B_2$ is the induced magnetic strength value of the second layer, and $F_{m1}$ is the interlayer magnetic interaction.

In a similar manner, we can deduce that:

$$B_i = B_{i-1} \cdot f(F_{m(i-1)}) \tag{34}$$

Where $B_i$ is the induced magnetic strength value of the $i$-th layer, and $F_{m(i-1)}$ is the interlayer magnetic interaction of the i-th layer.

Then we can obtain the induced magnetic strength of the $n$-th layer, where we measure the external magnetic strength, as follows:

$$B_n = B_0 \prod_{i=0}^{n-1} f(F_{mi}) \tag{35}$$

When the average interatomic distance between layers varies due to stress or fatigue, the interlayer magnetic interaction changes correspondingly by $\Delta F_{mi}$.

Then the induced magnetic field strength of the n-th layer can be expressed as:

$$B'_n = B_0 \prod_{i=0}^{n-1} f(F_{mi} + \Delta F_{mi}) \tag{36}$$

In the case where $\Delta F_{mi}$ is minimal, we have:

$$f(F_{mi} + \Delta F_{mi}) \approx f(F_{mi}) + f'(F_{mi}) \cdot \Delta F_{mi} \tag{37}$$

Where $f'(F_{mi})$ represents the derivative of function $f$ at $F_{mi}$. Substitute into $B'_n$ to obtain:

$$B'_n = B_0 \prod_{i=0}^{n-1} [f(F_{mi}) + f'(F_{mi}) \cdot \Delta F_{mi}] \tag{38}$$

By taking the logarithm of $B_n$ and $B'_n$ respectively, we can obtain:

$$\ln B_n = B_0 \sum_{i=0}^{n-1} \ln f(F_{mi}) \tag{39}$$

$$\ln B'_n = B_0 \sum_{i=0}^{n-1} \ln [f(F_{mi}) + f'(F_{mi}) \cdot \Delta F_{mi}] \approx B_0 \sum_{i=0}^{n-1} \ln \left[ f(F_{mi}) + \frac{f'(F_{mi}) \cdot \Delta F_{mi}}{f(F_{mi})} \right] \tag{40}$$

Then we have:

$$\ln B'_n - \ln B_n \approx B_0 \sum_{i=0}^{n-1} \frac{f'(F_{mi}) \cdot \Delta F_{mi}}{f(F_{mi})} \qquad (41)$$

Simplifying equation 44 yields:

$$\Delta B_n = B'_n - B_n \approx B_0 \left[ \exp\left( \sum_{i=0}^{n-1} \frac{f'(F_{mi}) \cdot \Delta F_{mi}}{f(F_{mi})} \right) - 1 \right] \qquad (42)$$

Where $\Delta B_n$ represents the change in magnetic strength that occurs when subjected to stress or fatigue.

Since when $x$ is minimal, $\exp(x) \approx 1 + x$. Equation 11 can be simplified as:

$$\Delta B_n \approx B_0 \left[ \sum_{i=0}^{n-1} \frac{f'(F_{mi}) \cdot \Delta F_{mi}}{f(F_{mi})} \right] \qquad (43)$$

Due to the correlation between $\Delta F_{mi}$ and the change in interlayer atomic spacing $\Delta R_i$ ($\Delta F_{mi} = K_m \Delta R_i$), we can obtain:

$$\Delta B_n \approx B_0 \sum_{i=0}^{n-1} \frac{f'(F_{mi}) \cdot K_m \Delta R_i}{f(F_{mi})} \qquad (44)$$

If the entire material is subjected to a uniform force or fatigue, the atomic spacing between each layer tends to remain consistent, and the magnetic interactions within each layer will also exhibit similar effects. To simplify the derivation, we assume that the variation in atomic spacing ($\Delta R_i$) between adjacent layers is equal, $\Delta R_i = \Delta \bar{R}$, furthermore, the magnetic interaction function $f(F_{mi})$ of each layer is also identical, that is, $f(F_{mi}) = f(F_m)$.

Equation 47 can be simplified as:

$$\Delta B_n \approx B_0 \cdot n \frac{f'(F_m) \cdot K_m \Delta \bar{R}}{f(F_m)} \qquad (45)$$

Note that using the same derivation, it can be deduced that the magnetic strength variation of any monolayer as:

$$\Delta B_i \approx B_{i-1} \cdot \frac{f'(F_{mi}) \cdot K_m \Delta R_i}{f(F_{mi})} \qquad (46)$$

Given that atomic spacing changes primarily occur within the PSB during the fatigue process of materials, we can consider that this magnification factor, $n$, corresponds precisely to the number of PSB layers. As fatigue progresses, the magnetic intensity changes induced by variations in the PSB within each layer will continuously accumulate and manifest macroscopically as a significant amplification of the magnetic intensity changes at the measurement points.

Due to the linear correlation when the interatomic distance $\Delta \bar{R}$ undergoes minimal changes, by substituting the relationship between interlayer bonding force and interlayer atomic spacing variation ($\Delta F_C = K_c \Delta \bar{R}$) into equation 48, we can obtain:

$$\Delta B_n = K_B \Delta F_C \qquad (47)$$

Where $K_B = \frac{B_0 n K_m f'(F_m)}{f(F_m) K_c}$

Because the magnetic strength $\Delta B_H$ measured by external Hall sensors is actually a mapping of the internal magnetic strength $\Delta B_n$, we obtain:

$$\Delta B_H = K_E \Delta B_n \tag{48}$$

Where $K_E$ is a constant in the case the measurement distance and environmental factors remain unchanged.

Then we can obtain:

$$\Delta B_H = K_H \Delta F_C \tag{49}$$

Where $K_H = K_E K_B$.

Consequently, we can observe the fatigue process of ferromagnetic materials, which is caused by the gradual accumulation and weakening of bonding forces, through the significant amplification effect of magnetic flux on subtle internal changes induced by external magnetic field strength.

**Static load test of Soft Magnetic Iron and waveform interpretation of magnetic strength variations**

The static load test on Soft Magnetic Iron reveals that the macroscopic magnetic strength of the quantum spin correlation varies with the loading displacement, mirroring the characteristic behavior of interatomic magnetic force changes with interatomic spacing. Both exhibit a peak point, with synchronous trends before the peak and opposing trends thereafter. This underscores a strong correlation between macroscopic magnetic strength changes and interatomic magnetic forces as atomic spacing varies. Notably, if the initial displacement of the load occurs before this peak point and the maximum displacement occurs after this peak point, the cyclic fatigue loading will result in an M-shaped waveform in the magnetic flux intensity.

**Fatigue quantum spin correlation testing of ferromagnetic materials under different cycling loads**

We conducted fatigue quantum spin correlation testing on Soft Magnetic Iron, three types of steel: 45 Steel (C45E4, widely used in engine components, transmission systems, suspension systems, and building structures), Q235 Steel (PB240, widely used in high-voltage transmission towers, mechanical manufacturing, and building structures), Bridge Stay Cable Steel (BSCS) and Pure Nickel (Nickel 201) (Extended Data Fig. 2).

We discovered that the greater the average rate of change in MagDrift, which we denote as $k_{av}$ and corresponds to the slope of the linear fitting line for MagDrift, the fewer cycles were observed. Prior to fracture, $k_{av}$ and the number of cycles exhibited a nearly linear correlation. Upon normalizing the fatigue life across all MagDrift curves, it becomes evident that the normalized $k_{av}$ are highly consistent. (The presence of a larger fatigue zone, Extended Data Fig. 1b, for test specimen under high stress conditions leads to a slightly greater total change in fatigue-induced MagDrift at fracture compared to lower stress levels). This suggests that the fatigue behavior of the test specimens adheres to similar variation patterns, and the Change Rate of Magnetic Strength (CRMS) (Extended Data Fig. 3a-d) attributed to fatigue is equivalent to the rate of weakening in bonding force, ultimately determining the ultimate fatigue life.

**Static load testing of ferromagnetic materials**

Integrating the static load data of five materials (Extended Data Fig. 3d), and discovering that the static magnetic strength changes of 45 Steel, Q235 Steel, and Soft Magnetic Iron all have a process of increasing and then decreasing with the loading displacement in the initial stage, hence the fatigue MagDrift curves of these three materials have similar trends when they are about to fracture. The magnetic strength of Bridge Stay Cable Steel decreases first with displacement and then increases in the final stage. The magnetic strength of Pure Nickel only monotonically decreases. Therefore, the fatigue MagDrift curves of these two materials show different trends when about to fracture compared to Soft Magnetic Iron. The reason for this phenomenon may be lie in the different atomic spacings of different metals.

**Fatigue testing under typical cyclic loading conditions**

We selected three typical cyclic stress states for analyzing the changes in strain, displacement, and magnetic strength under fatigue loading conditions.

Initially, there is a slight increase in mean strain, followed by minimal variation throughout most of the cycle, culminating in a rapid increase immediately prior to fracture. Regardless of cyclic stress levels, the strain waveform remains consistent, with peak-to-peak strain variation remaining virtually constant until a notable deviation occurs just before fatigue fracture. The strain amplitude changes under the three stress conditions are approximately 0.42%, 0.43%, 0.43%, respectively. The proportions of strain change from the beginning to near fracture under the three stress conditions are approximately 2.17%, 1.01%, 0.42%, respectively.

In contrast to the subtle strain changes during fatigue, the quantum spin correlation magnetic strength and peak-to-peak values under cyclic loading exhibit pronounced variations. Magnetic strength decreases rapidly at the onset of loading and continues to decrease gradually as cyclic loading progresses. Notably, a rapid decrease in magnetic strength precedes fatigue fracture, significantly earlier than the onset of rapid strain changes. The rate of magnetic strength change prior to fracture increases significantly compared to the initial rate. The proportions of magnetic strength amplitude changes under the three stress conditions are 31.46%, 32.37%, and 27.28%, respectively.

To isolate magnetic changes due to fatigue from those caused by loading, we define MagDrift as the amplitude change between the valley point 'c' and peak point 'd'. Setting the initial MagDrift at the start of loading as the fatigue-free state, changes in MagDrift from loading onset to fatigue fracture reflect magnetic flux intensity variations during fatigue. (Theoretically, zero-stress magnetic strength could also indicate fatigue-induced changes, but achieving true zero stress at 20Hz loading frequency is impractical due to machine vibrations.)

It shows that MagDrift undergoes an abrupt change as fracture approaches, significantly earlier and larger than displacement amplitude variation. The variation ratios of MagDrift under the three cyclic loading forces are 22.45%, 12.86%, and 9.67%, respectively.

Comparing magnetic strength amplitude changes due to loading to strain amplitude changes, and MagDrift to material elongation due to fatigue across 19 specimen tests, we found that magnetic strength changes due to loading are 28.61 to 86.72 times greater than strain changes. Similarly, the rate of MagDrift change due to fatigue is 8.05 to 40.20 times greater than the rate of material elongation caused by fatigue (Fig. 4).

Evidently, magnetic strength exhibits significantly higher sensitivity to fatigue changes than mechanical variables.

**Fatigue life evaluation method.**

To evaluate the final fatigue life using early cyclic loading data, we devised an analysis method, taking Soft Magnetic Iron as an illustrative example.

Firstly, the quantum spin correlation detection method was employed to conduct fatigue experiments on 19 specimens under varying stress levels, during which magnetic strength was obtained throughout the fatigue process. The MagDrift curve, representing the maximum peak-to-peak magnetic intensity per second loading cycle, was calculated. A linear fitting curve was then fitted to the average rate of change of MagDrift in order to determine the $k_{av}$ values. Subsequently, the fatigue failure lives and their corresponding $k_{av}$ values, obtained at different stress levels, were plotted as M-N curves. From these curves, the M-N linear fitting equation was derived as:

$$\log|k_{av}| = k_c \log N + C \tag{50}$$

Then we obtain:

$$N = 10^{\frac{1}{k_c}(\log|k_{av}|+C)} \tag{51}$$

Where $k_c$=-1.261, $C$=-0.167.

Substitute $k_{av}$ obtained from the MagDrift for the early loading cycles (5%, 6%, 7%, 8%, 9%, 10%, 20%, 30%, 40%, 50%, 60% of the fatigue failure life) of the 19 specimens into the aforementioned equation to calculate their respective fatigue failure lives. Subsequently, the calculated fatigue failure lives were correlated with the actual fatigue failure lives of the 19 specimens using the following equation.

$$r = \frac{\Sigma(x-\bar{x})(y-\bar{y})}{\sqrt{\Sigma(x-\bar{x})^2 \Sigma(y-\bar{y})^2}} \tag{52}$$

Where $x$ is the calculated fatigue failure life and $y$ is the actual fatigue failure life.

It can be calculated that the correlation between the calculated fatigue life and actual life before 5%, 6%, 7%, 8%, 9%, 10%, 20%, 30%, 40%, 50%, and 60% of the loading cycle is 0.737, 0.762, 0.834, 0.886, 0.907, 0.911, 0.961, 0.977, 0.986, 0.993, and 0.994, respectively.

**Fatigue fracture early warning algorithm.**

We propose a two windows algorithm. When the number of cycles does not exceed 300,000, starting from the 5,000th cycle loading, the first window covers the 60 cycles immediately preceding the current time, and the second window covers the 1,000 cycles preceding the point 3,000 cycles before the current time plus an additional 60 cycles. Let $B_{med}$ represent the median MagDrift value within the first window, and $A_{med}$, $A_{max}$, and $A_{min}$ represent the median, maximum, and minimum MagDrift values within the second window. The alarm threshold is set as $Alarm_{th} = B_{med} - A_{med} + \frac{A_{max}-A_{min}}{S_1} < 0$, where $S_1$ is a constant obtained through experimentation. For Soft Magnetic Iron, the value of $S_1$ is 4.8 and the value for $A_{max} - A_{min}$ is 0.0132. When the alarm threshold falls below 0 for more than four consecutive measurements, the alarm is triggered. Once the number of cycles exceeds 300,000, the first window covers the 60 cycles immediately preceding the current time, and the second window covers the 2,000 cycles preceding the point 40,000 cycles before the current time plus an additional 60 cycles. Set the alarm threshold as $Alarm_{th} = B_{med} - A_{med} - \frac{A_{max} - A_{min}}{S_2} > 0$, where $S_2$ is also a constant obtained through experimentation. For Soft Magnetic Iron, the value of $S_2$ is 1.2 and the value for $A_{max} - A_{min}$ is 0.0132. When the alarm threshold exceeds 0 for more than four consecutive measurements, the alarm is triggered.

## Data availability

All data are available within the main text and Supplementary Materials. Access to the original datasets used in this study requires formal request through the corresponding author via email: benniuzhang@cqjtu.edu.cn

**Acknowledgments** We thank H. Li for fruitful discussion. We thank Chongqing Wanqiao Traffic Technology Development Limited Company for providing Bridge Stay Cable Steel material. This research is supported by National Natural Science Foundation of China (grant number 52178272 and 52478300). We gratefully acknowledge support from Sichuan Tibetan Area Expressway Co., Ltd (2021-03) and China Railway Construction Kunlun Investment Group Co., Ltd (YYGL-XZ-2024-008).

**Author contributions** B.N.Z. conceived the study, derived the theoretical formula, designed the experiment, analyzed the experimental data, and wrote the manuscript. L.S.Z. performed the experiments, analyzed the experimental data, plotted figures and wrote the manuscript. X.D.W. derived the theoretical formula, performed the experiments, analyzed the experimental data, plotted figures and wrote the manuscript. G.J.Y. performed the Electron Probe Micro-Analysis tests, plotted some figures and wrote the manuscript. X.C.C. designed the experiment and analyzed the experimental data. Z.J.Z. Participate in the experiments, search literatures and plotted some figures. X.L. Participate in the experiments and plotted some figures. F.P.Z. Participate in the experiments, analyzed the experimental data and plotted some figures. J.L.P. Participate in the experiments and search literatures. H.F.J. wrote and revised the manuscript. G.Z. revised the manuscript. All authors contributed to the preparation of the manuscript.

**Competing interests** B.N.Z., J.G.Y., and L.S.Z. have filed a patent at CNIPA (CN Application no. 202410434371.4). B.N.Z., and Z.J.Z. have filed a patent at CNIPA (CN Application no. 202411680900.5). B.N.Z., X.L., F.P.Z., L.S.Z., and X.D.W. have filed a patent at CNIPA (CN